\documentclass[galaxies,article,accept,moreauthors,pdftex]{mdpi} 
\firstpage{1} 
\pubvolume{1}
\issuenum{1}
\articlenumber{0}
\pubyear{2023}
\copyrightyear{2023}
\externaleditor{Academic Editor: Jorick Vink}
\datereceived{24 April 2023} 
\daterevised{31 May 2023} 
\dateaccepted{1 June 2023} 
\datepublished{} 
\hreflink{https://doi.org/} 
\usepackage{subcaption}
\usepackage{tabularx}
\Title{New Insight into the FS\,CMa System MWC 645 from Near-Infrared and Optical Spectroscopy}

\TitleCitation{New Insight into the FS\,CMa System MWC 645 from Near-Infrared and Optical Spectroscopy}

\Author{Andrea Fabiana Torres $^{1,2,}$*\orcidA{}, Mar\'ia Laura Arias $^{1,2}$\orcidB{}, Michaela Kraus $^{3}$\orcidC{}, Lorena Ver\'onica Mercanti $^{1,2}$ and T\~{o}nis Eenm\"ae $^{4}$\orcidD{}}

\AuthorNames{Andrea F. Torres, Mar\'ia L. Arias, Michaela Kraus, Lorena V. Mercanti and T\~{o}nis Eenm\"ae}

\AuthorCitation{Torres, A.F.; Arias, M.L.; Kraus, M.; Mercanti, L.V.; Eenm\"ae, T.}

\address{%
$^{1}$ \quad Departamento de Espectroscop\'ia, Facultad de Ciencias Astron\'omicas y Geof\'isicas, Universidad Nacional de La Plata, Paseo del Bosque S/N, \mbox{B1900FWA La Plata, Argentina}\\
$^{2}$ \quad  Instituto de Astrof\'isica de La Plata (CCT La Plata-CONICET, UNLP), Paseo del Bosque S/N, \mbox{B1900FWA La Plata, Argentina}\\
$^{3}$ \quad Astronomical Institute, Czech Academy of Sciences, Fri\v{c}ova 298, \mbox{251\,65 Ond\v{r}ejov, Czech Republic}
\\
$^{4}$ \quad Tartu Observatory, University of Tartu, Observatooriumi 1, 61602 T\~{o}ravere, Estonia}

\corres{Correspondence: atorres@fcaglp.unlp.edu.ar}

\abstract{The B[e] phenomenon is manifested by a heterogeneous group of stars surrounded by gaseous and dusty circumstellar envelopes with similar physical conditions. Among these stars, the FS\,CMa-type objects are suspected to be binary systems, which could be experiencing or have undergone a mass-transfer process that could explain the large amount of material surrounding them. We aim to contribute to the knowledge of a recently confirmed binary, MWC\,645, which could be undergoing an active mass-transfer process. We present near-infrared and optical spectra, identify atomic and molecular spectral features, and derive different quantitative properties of line profiles. Based on publicly available photometric data, we search for periodicity in the light curve and model the spectral energy distribution. We have detected molecular bands of CO in absorption at 1.62 $\upmu$m and 2.3 $\upmu$m for the first time. We derive an upper limit for the effective temperature of the cool binary component. We found a correlation between the enhancement of the H$\alpha$ emission and the decrease in optical brightness that could be associated with mass-ejection events or an increase in mass loss. We outline the global properties of the envelope, possibly responsible for brightness variations due to a variable extinction, and briefly speculate on different possible scenarios. }

\begin{document}



\section{Introduction}\label{intro}

In their evolution, some B-type stars undergo phases that are still puzzling for astrophysicists even after years of study since they develop certain peculiarities that are not yet well understood. The B[e] phenomenon displayed by several B-type stars in their optical spectra is an example of them. Its manifestation can be seen through the presence of permitted and forbidden low-excitation emission lines of neutral and low ionization metals arising from circumstellar (CS) gas and large infrared excess due to CS dust \citep{1976A&A....47..293A}. The phenomenon is associated with stars with different initial masses, isolated or in binary systems, transiting different evolutionary stages, such as supergiants, compact planetary nebulae, Herbig Ae/Be stars, and symbiotic systems \citep{1998A&A...340..117L}. Despite the cited differences among the stars, the physical conditions of their CS gaseous and dusty envelopes are similar, which is a crucial factor when seeking to understand the development of the phenomenon. Furthermore, since the CS envelopes veil the photospheric features of the central objects, it is difficult to determine their spectral types and evolutionary states. Therefore, stars without a proper classification comprise the category named ``Unclassified B[e] stars'' (UnclB[e]).

Nearly a decade after Lamers' classification, Miroshnichenko proposed a new group called FS\,CMa stars \citep{2007ApJ...667..497M}. The observational defining criteria are 1) the presence of a hot star (between O9 and A2 spectral types) continuum with emission lines of H I, Fe II, O I, [Fe II], [O I], Ca II; 2) an infrared spectral energy distribution (SED) that shows a large excess with a maximum at 10--30 $\upmu$m and a strong decrement beyond these wavelengths; and 3) a star location outside a region of star formation. According to these properties, almost all the UnclB[e] objects in Lamers et al.'s publication are in this group \citep{2022A&A...659A..35K}, which has approximately seventy members between confirmed and candidate ones \citep{2019OAP....32...63K}. They are suspected to be binaries at a post-mass-exchange evolutionary phase, with a secondary component fainter and cooler than the primary or degenerate \citep{2007ApJ...667..497M, 2007ApJ...671..828M, 2009ApJ...700..209M}. Since the predictions of mass-loss rates of single-stars theory \citep{2001A&A...369..574V, 2010ApJ...721.1079C} cannot explain the existence of a large amount of CS matter, a mass-transfer process in a binary system could be a likely explanation. However, only a few objects of this class have been confirmed as binary systems, probably due to the scarcity of available observational data to discover them and the difficulties in detecting signs of binarity due to the presence of CS matter, the intrinsic stellar variability, and the low brightness of most members of the FS\,CMa group \citep{2020MmSAI..91..118K}. \mbox{Recently, \citet{2023Galax..11...36M}} published a review of FS\,CMa objects, where they reported fifteen stars as binaries and six as binary system candidates.

MWC\,645 (= V2211\,Cyg, $\alpha$ = 21:53:27.49, $\delta$ = +52:59:58.01; V = 13.0, H-K = 1.53, \linebreak J-H = 1.267) was originally included in the supplement of the Catalogue of Mount Wilson about A and B stars with bright H I spectral lines \citep{1943CMWCI.682....1M}. The presence of strong double emission lines of Fe II and [Fe II] (with a radial velocity difference between the red and blue peaks of 150 km s$^{-1}$) and triple-peaked profiles of the H$\gamma$ and H$\delta$ transitions were reported by \citet{1973ApL....14...65S}. They also remarked striking spectral similarities between MWC\,645 and $\eta$\,Car. Also, permitted and forbidden transitions of low excitation and ionization potential belonging to Ti II, Cr II, [O I], and [N II] were observed \citep{1976A&A....47..293A}. Photometric variations were found by \citet{1978ApJ...225..488G} with an amplitude of 0.3\,mag and a possible period of 23.6 years. A deep spectroscopic study was done by \citet{1996A&AS..120...99J} that revealed no stellar absorption features. These authors concluded that possibly MWC\,645 is a late B-type object based on the absence of He II lines and the weakness of the He I lines at $\lambda$ 6678 \AA\, and $\lambda$ 7065 \AA\, possibly detected once, each one at different years. They did not find spectral transitions from C, Ne, and Mg atoms or ions, but they found lines of K I and Cu II (typically seen in stellar types later than F) and Zr II (usually seen in stars later than A0-type). They highlighted the extreme spectroscopic variability of MWC\,645 over the years. \citet{1998A&A...340..117L} included it in the UnclB[e] stars group. MWC\,645 has IRAS flux ratios that locate it in the region occupied by OH/IR stars \citep{2001A&A...376..917L}. \citet{2003A&A...408..257Z} detected a characteristic asymmetric profile for the emission metal lines, with a steep red flank and a blue wing. He reported the splitting in the central emission of [O I] and [Fe II] lines and a peculiar emission profile of the H$\alpha$ line showing a broad blue and a narrow red component with a full width at half maximum (FWHM) of 5.0 \AA\, and 1.3 \AA\,, respectively. He proposed a latitude-dependent wind model with a large optical depth dust disk at an intermediate inclination to explain the asymmetric line profiles and their splitting. \citet{2008A&A...477..193M} obtained H$\alpha$ narrow band imaging and found no visible extended emission associated with the star.

Recently, \citet{2021OAP....34...59N, 2022ApJ...936..129N} studied high-resolution optical spectra of MWC\,645 taken in two different years. He found absorption lines of neutral metals, such as Li I, Ca I, Fe I, Ti I, V I, and Ni I, typically present in cool stellar spectra, with a different average radial velocity in each spectrum, that revealed the binary nature of the object. However, they did not find any absorption line typical of a B-type object in any of their spectra. They disentangled the contribution of each stellar component and estimated their surface temperatures and luminosities (T$_{eff}$ = 18\,000 $\pm$ 2000 K and 4250 $\pm$ 250 K, \linebreak log (L/L$_{\odot}$) = 4.0 $\pm$ 0.5 and 3.1 $\pm$ 0.3 for the hot and cool components, respectively). Low-resolution near-IR spectra displayed emission lines of the H I Paschen and Brackett series, as well as of Fe II, O I, N I, and He I. Photometric monitoring in the optical and near-IR regions showed quasi-cyclic variations of both short and long periods (months \mbox{and $\sim$ 4 years,} respectively). The authors conclude that the star can be classified as an FS\,CMa-type object, where its intermediate-mass components (7 M$_{\odot}$ and 2.8 M$_{\odot}$) undergo an ongoing mass-transfer process. According to the shape displayed by the spectral energy distribution with weak emission peaks at about 10 $\upmu$m and 18 $\upmu$m, they inferred the presence of silicates in an optically thin dusty shell.

MWC\,645 is one of the eight FS\,CMa objects in which absorption lines of neutral metals typical of late-type secondaries have been detected. To contribute to the study of this intriguing object, we decided to observe it in the near-IR to search for signatures of both stars, mainly of the cool component, that help to characterize it. In addition, the acquisition of new optical spectra and the public availability of data (spectroscopic and photometric) that could shed some light on this complex system motivated us to analyze them. \mbox{The paper} is organized as follows: We present the infrared and optical observations used in this work in Section \ref{observations}. In Sections \ref{ir} and \ref{optical}, we analyze the data. In Section \ref{discussion}, we discuss the results. Finally, Section \ref{conclusions} contains the main conclusions.

\section{Observations}\label{observations}

\subsection{Near-Infrared Spectra}

Near-infrared spectra were taken using the Gemini Near-Infrared Spectrograph (GNIRS, \citep{2006SPIE.6269E..4CE}) attached to the 8 m telescope at GEMINI-North (Hawaii) under the programs GN-2017A-Q-62, GN-2018A-Q-406, and GN-2022B-Q-225. On 6 June 2017, we obtained $K$-band spectra in long-slit mode centered at 2.35 $\upmu$m. The instrumental configuration used was a 110.5 l/mm grating, a 0.3 arcsec slit, and the short camera (0.15 arcsec/pix). We also acquired spectra with the same configuration but in cross-dispersed mode centered at \linebreak 2.19 $\upmu$m and 2.36 $\upmu$m on 24 and 30 July 2018, respectively. The effective spectral coverage by these set-ups was 0.90 $\upmu$m--2.27$\upmu$m and 0.85 $\upmu$m--2.45 $\upmu$m, respectively, with gaps between the orders (the interval 1.36 $\upmu$m to 1.46 $\upmu$m is unusable due to saturated telluric lines). \mbox{The resulting} mean spectral resolving power of the spectra was R$\sim$5500. On 24 August 2022, $L$-band spectra were obtained with a different long-slit configuration: a 31.7 l/mm grating, a 0.1 arcsec slit, and the long camera (0.05 arcsec/pix), with two different central wavelengths (3.48 and 4.00 $\upmu$m). This configuration resulted in R$\sim$5100.

The spectra were taken in two ABBA nodding sequences along the slit. To account for telluric absorption, a late-B- or an early-A-type star close to the target in both time and position was observed. Stars of these spectral types are featureless in the observed wavelength range, except for hydrogen absorption lines that can be successfully removed in the reduction process by fitting theoretical line profiles. Flats were also acquired. The data were reduced with the Image Reduction and Analysis Facility (IRAF)/Gemini tasks. \mbox{The sky} contribution was removed by subtracting the AB pairs. The spectra were flat-fielded and telluric corrected. The wavelength calibration was performed using the telluric lines. The data were normalized to unity.

A and B positions were added to increase the signal-to-noise ratio (S/N). The final S/N ratio varies for the different spectral ranges, as it is affected by the quality of the telluric correction. Some regions are very polluted with telluric lines and it was impossible to make a complete cancellation, thus some residuals remain. In addition, for some spectral regions heavily crowded by emission lines, it becomes difficult to make accurate S/N ratio estimates. Table \ref{signal-noise} summarizes the mean values of the S/N ratio for all our GNIRS \mbox{near-IR observations.}

\begin{table}[H]
\caption{Mean values of the S/N ratio for GNIRS near-IR observations in different spectral ranges.}
\begin{tabularx}{\textwidth}{p{6cm}rr}
\toprule
\textbf{Observations}&\textbf{Spectral  Range} & \textbf{Mean S/N Ratio}\\
\textbf{Program ID}          & \textbf{[\AA]}   &        \\  
\midrule
GN-2017A-Q-62 & 22,570--24,400 & 200 \\
GN-2018A-Q-406&21,000--24,500 & 200\\
GN-2018A-Q-406&15,800--18,200& 100 \\
GN-2018A-Q-406 & 11,000--13,600 & 90   \\
GN-2018A-Q-406 & 8400--11,000 &  60\\
GN-2022B-Q-225 &  33,311--36,446 & 200\\
GN-2022B-Q-225 &   38,500--41,786& 100\\ 
\bottomrule
\end{tabularx}
\label{signal-noise}
\end{table}


\subsection{Complementary Data}

Optical observations were carried out at Ond\v{r}ejov Observatory, Czech Republic, using the Coud\'e spectrograph \citep{2002PAICz..90....1S} attached to the Perek 2 m telescope. We obtained spectra with a resolving power of R$\sim$12\,000 covering a spectral range from 6262 \AA\, to 6735 \AA\, on 12 and 13 September 2018. We also acquired a spectrum centered at 8600 \AA. We used a grating of 830.77 l/mm with a SITe 2030 $\times$ 800 CCD and a slit width of 0.7 arcsec. Additional observations were done at Tartu Observatory, Estonia, with the 1.5 m Cassegrain reflector AZT-12 on 1 November 2021, using the long-slit spectrograph ASP-32 with a 600 l/mm. The wavelength coverage extended from 5450 \AA\, to 7480 \AA. Data were processed using standard IRAF tasks. Spectra were bias and flat-field corrected, wavelength calibrated, heliocentric velocity corrected, and flux normalized. 

\textls[-15]{We also searched for available optical spectra in the BeSS database \citep{2011AJ....142..149N}. We downloaded sixteen spectra taken between 2019 and 2022, with a resolving power R $\sim$ 14,000/16,000} in the spectral range 6500--6600 \AA\,. In addition, we collected a lower resolution spectrum (R $\sim$ 5000) acquired on 30 August 2019 that covers the range 6150--7000 \AA\,. The spectra were corrected by heliocentric velocity and normalized to the continuum using the standard IRAF tasks. We chose the same sample of continuum points to normalize all spectra. \mbox{The telluric} correction has not been applied.

In addition, we extracted from the public database ASAS-SN (All-Sky Automated Survey for Supernovae; \citet{2014AAS...22323603S, 2017PASP..129j4502K}) (\url{https://www.astronomy.ohio-state.edu/asassn/}(accessed on 2 February 2023)) survey photometric data of this star obtained over eight years. The collection is composed of {\it V}-band magnitudes from 16 December 2014 to 29 November 2018 and data in the {\it g}-band from 12 April 2018 up to 17 January 2023. 
Furthermore, we collected ground- and spaced-based multicolor photometry from Vizier service from 0.3 $\upmu$m to 140 $\upmu$m.

\section{Analysis of the IR Data}\label{ir}

\begin{figure}[H]
  \includegraphics[angle=0,width=1.\textwidth]{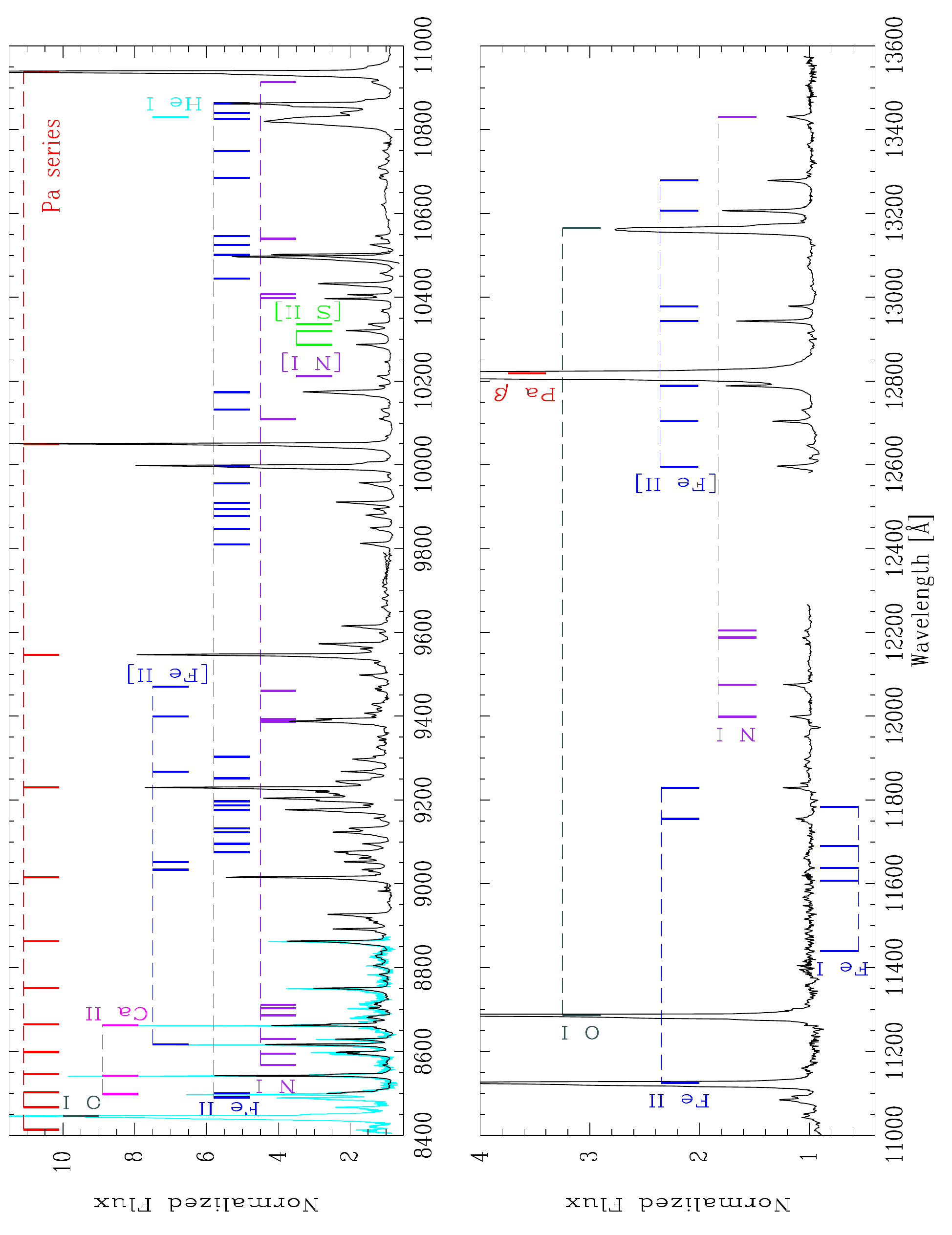}
  \caption{Normalized medium-resolution spectrum of MWC\,645 taken with Gemini/GNIRS on July 2018 from 8500 \AA\, to 13,600 \AA. The normalized Ond\v{r}ejov spectrum from 8400 \AA\, to 8870 \AA\,, acquired on September 2018, is shown in cyan. Main spectral lines are identified by colored markings. The spectral features of a given element (either permitted or forbidden and of different ionization states) are joined by a dashed line of the same color: hydrogen is indicated in red, oxygen in gray, iron in blue, calcium in pink, nitrogen in violet, sulfur in green, and helium in cyan. Wavelengths are given in angstroms.}
  \label{Figure-lines-8400-13600-mwc645} 
\end{figure}

Figure \ref{Figure-lines-8400-13600-mwc645} shows the near-IR spectrum of MWC\,645 from 8400 to 13,600 \AA. It displays numerous emission lines, particularly of the H I Paschen series. The strongest lines, except those of H I, correspond to O I, Fe II, and the Ca II triplet. Many transitions of N I can be identified in emission along this spectral range. Forbidden lines of [Fe II] and [S II] are also present. The moderate-resolution data reveal several absorption lines of Fe I that could be associated with the cool stellar companion. 

We carefully searched for He I lines in our spectra. If the lines are present, they are incipient and hidden in the noise. The most intense transitions in the interval $\lambda$ 8400--24,500 \AA\, (cited in the NIST database \citep{NIST_ASD}) correspond to $\lambda$10,830 \AA\, and $\lambda$20,587 \AA. \mbox{\citet{2022ApJ...936..129N} reported} the presence of the He I $\lambda$10,830 \AA\, transition in emission from a low-resolution spectrum (R $\sim$ 700). We identified a group of Fe II lines with the three emission peaks at the interval 10,826--10,862 $\upmu$m. However, the bluest feature of this group (see Figure \ref{Figure-lines-8400-13600-mwc645}) is broad, and thus, the He I $\lambda$10,830 \AA\, line could be blended with the Fe II lines. Our data have a higher resolution than the spectrum of Nodyarov et al., but not sufficient to separate the Fe II lines from the He I line. These authors also identified the He I $\lambda$20,587 \AA\, line. Unfortunately, it lies outside our spectral coverage. 

\begin{figure}[H]   
  \includegraphics[angle=0,width=1.\textwidth]{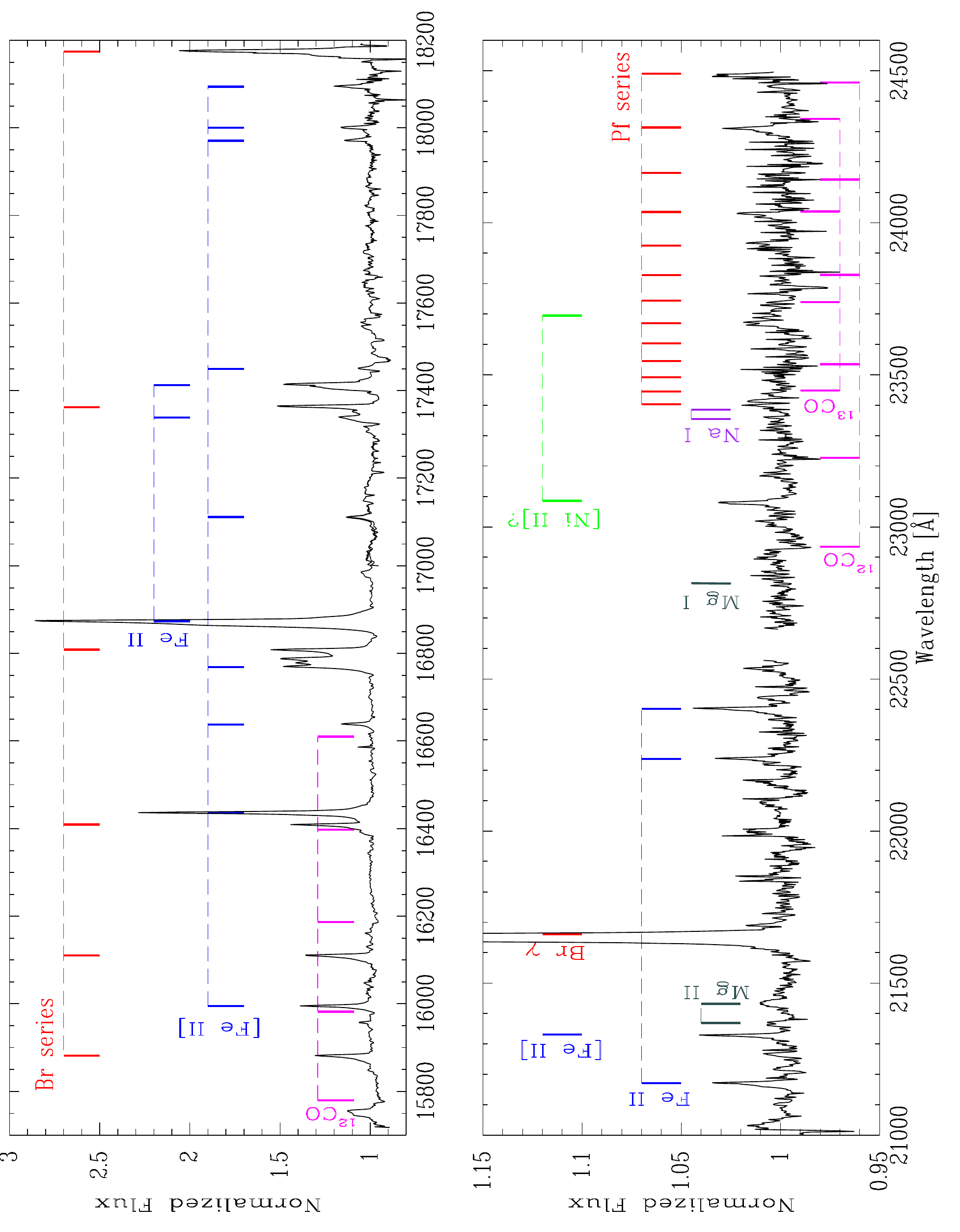}
  \caption{Normalized medium-resolution spectrum of MWC\,645 taken with Gemini/GNIRS in 2018, covering the $H$-(upper panel) and $K$-bands (lower panel). Main spectral lines and molecular bands are identified by colored markings. The spectral features of a given element (either permitted or forbidden and of different ionization states) or molecule (of different isotopes) are joined by a dashed line of the same color: hydrogen is indicated in red, magnesium in gray, iron in blue, sodium in violet, nitrogen in green, and carbon monoxide in pink. Wavelengths are given in angstroms.}
  \label{Figure-lines-15800-24500-mwc645} 
\end{figure}

The $H$-band spectrum of MWC\,645 (upper panel of Figure \ref{Figure-lines-15800-24500-mwc645}) is dominated by the H I Brackett series and several permitted and forbidden lines of Fe II. In the $K$-band, the most intense feature is the Br$\gamma$ line (see lower panel of Figure \ref{Figure-lines-15800-24500-mwc645}), which stands out among several emission lines corresponding to Fe II, [Fe II], and presumably [Ni II]. The Mg II doublet at $\lambda\lambda$ 21,374 \AA\, and 21,437 \AA\, is also in emission. The Pfund series extends from 2.3 microns longward. In addition, absorption features of neutral metals characteristic of late-type stars, such as Ca I, Mg I, and Na I, are present. For the first time, we have detected the presence of CO band heads in absorption around 2.3 $\upmu$m and around 1.6 $\upmu$m, which are typical photospheric features of late-type luminous stars.
\vspace{-3pt}

\begin{figure}[H]   
  \includegraphics[angle=270,width=1.\textwidth]{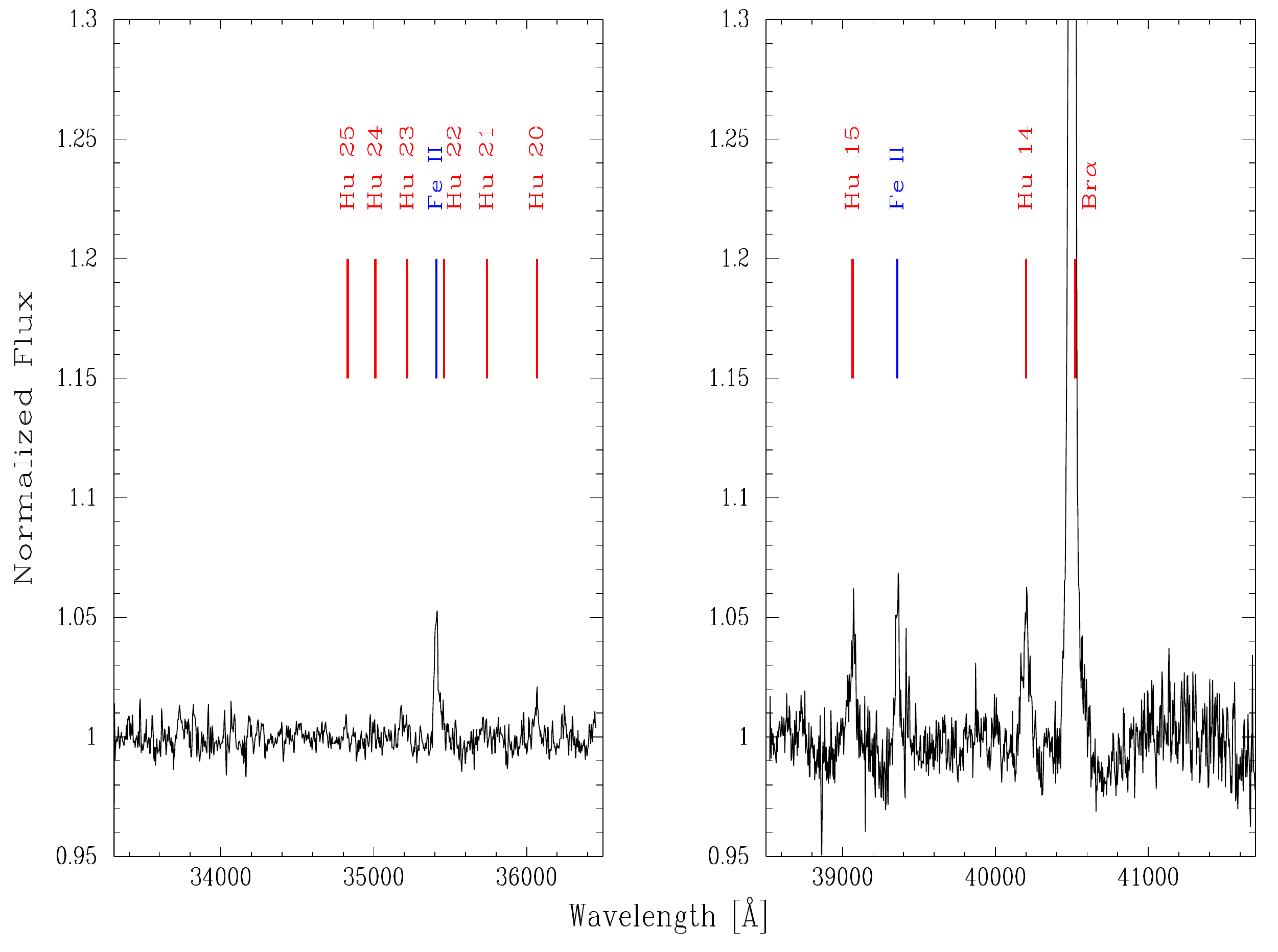}
  \caption{Normalized $L$-band spectrum of MWC\,645 obtained in 2022. The emission lines of H I and Fe II are marked in red and blue, respectively. The ``bump'' longward of the Br$\alpha$ line is a remnant from telluric correction. Wavelengths are in angstroms.}
  \label{fig-mwc645-banda-l} 
\end{figure}

Figure \ref{fig-mwc645-banda-l} shows the first obtained $L$-band spectrum of MWC\,645 in two different spectral regions. The first interval between 33,310 \AA\, and 36,410 \AA\, is relatively featureless (see left panel), except for the presence of the permitted emission line of Fe II $\lambda$ 35,423 \AA\,, which is clearly seen above the continuum level, and some H I lines of the Humphreys series in emission, where the strongest is the one corresponding to the 20-6 transition. Lower-order members of the Humphreys series can be seen in the second spectral interval (right panel) that ranges from 38,520 \AA\, to 41,730 \AA, where the strong emission of the Br$\alpha$ line can also be observed. We searched for absorption bands of the first-overtone of silicon monoxide (SiO) around 4 $\upmu$m but found none.

\begin{figure}[H]
  \includegraphics[angle=270,width=1.\textwidth]{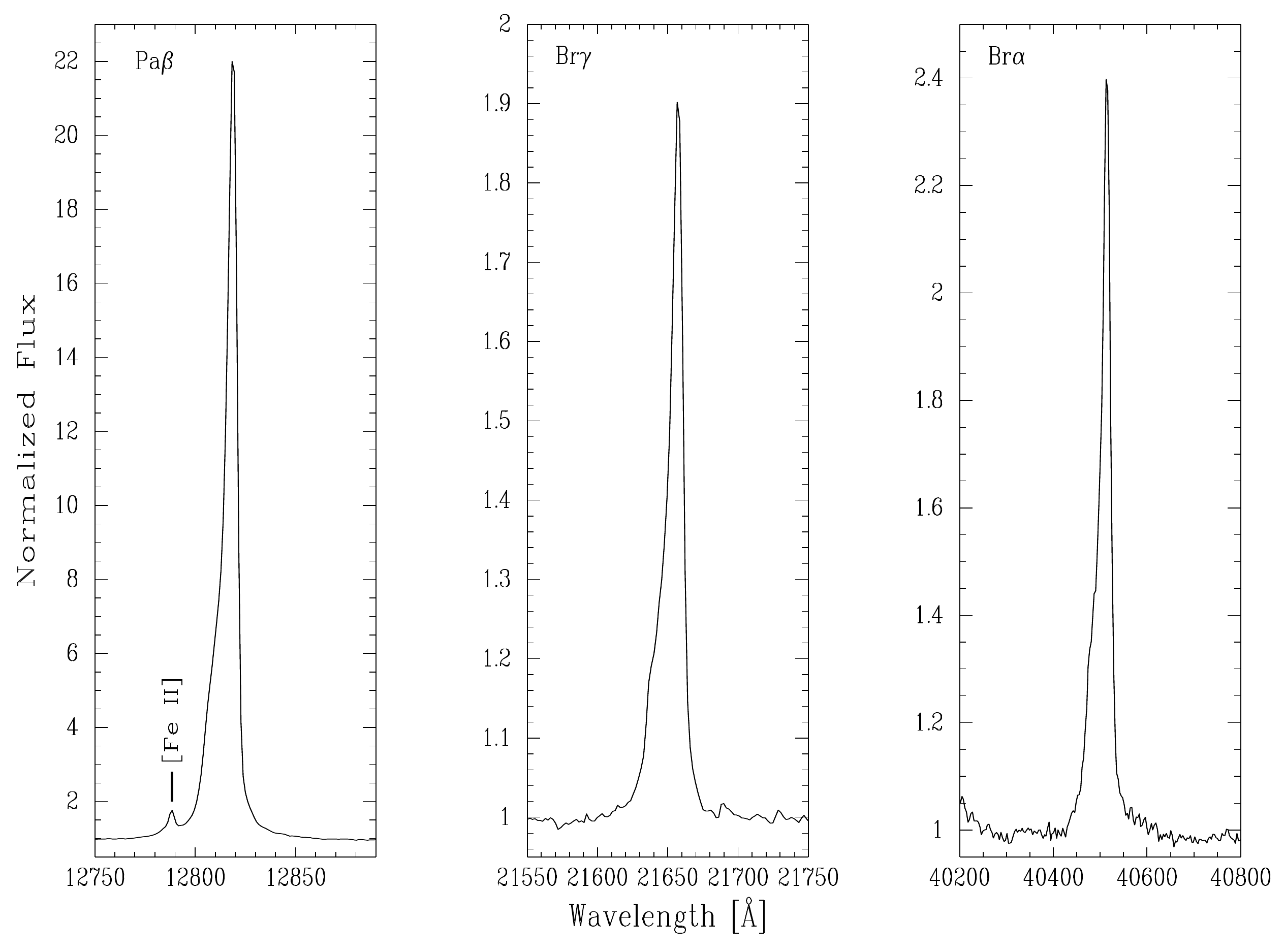}
  \caption{Strongest H I lines detected in our IR spectra of MWC\,645. They display an asymmetric profile, where the red flank is steeper than the blue one. Wavelengths are in angstroms.}
  \label{fig-mwc645-hi-lines} 
\end{figure}

Figure \ref{fig-mwc645-hi-lines} plots three H I lines: Pa$\beta$, Br$\gamma$, and Br$\alpha$. Their profiles are single-peaked but asymmetric. We measured the total equivalent width (EW) of each line using the 'e' function in the IRAF splot routine. The total measured EWs are 184 \AA\,, 15 \AA\,, and 48 \AA\,, respectively. The percentage uncertainty of the EW measurements is 5\%. Unfortunately, the wavelength calibration of our near-IR spectra is not accurate enough to determine reliable radial velocity measurements. 

\noindent{\em CO Absorption Bands}
\vspace{4pt}

\begin{figure}[H]   
  \includegraphics[angle=270,width=1.0\textwidth]{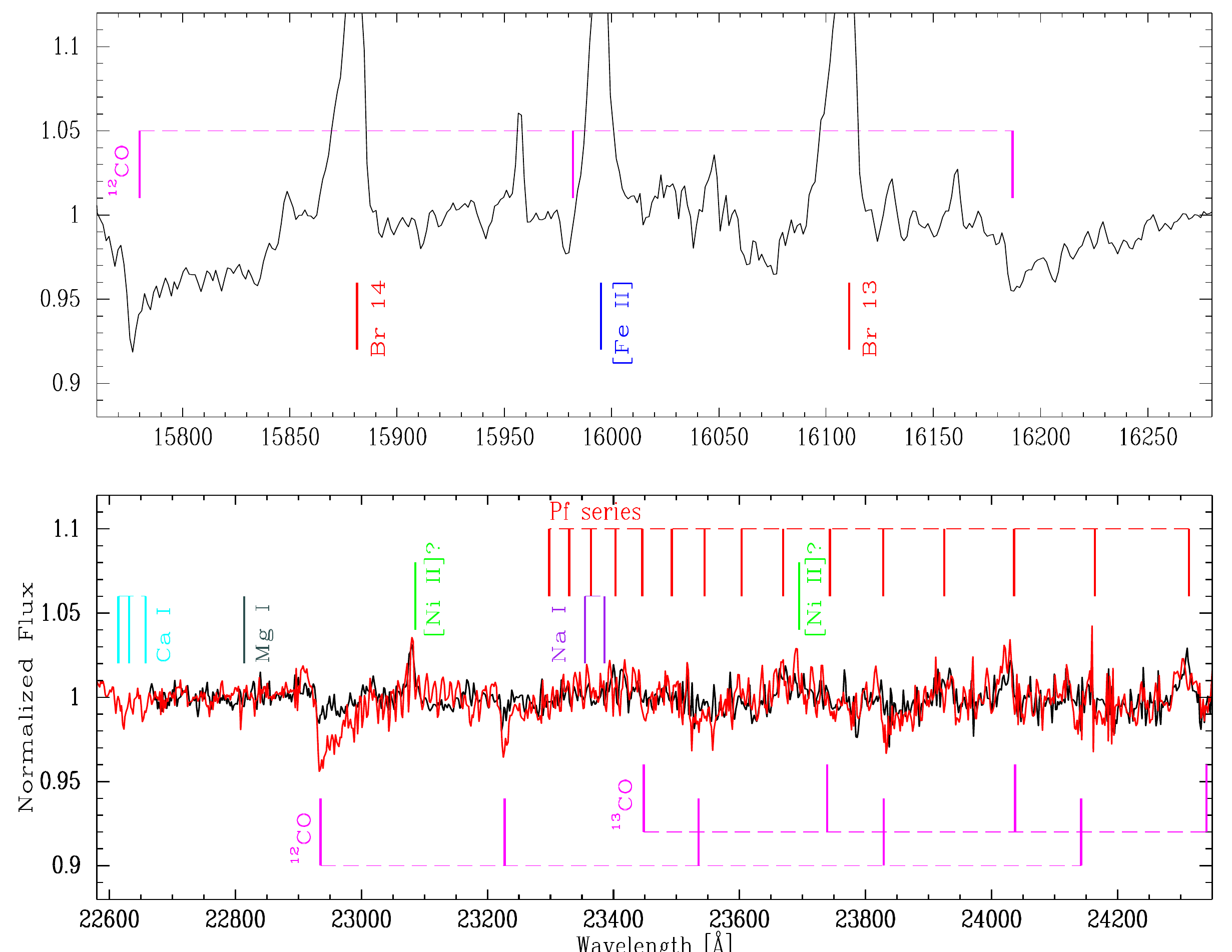} 
  \caption{CO molecular bands of MWC\,645. Some emission and absorption lines are also identified. The spectral features of a given element or molecule (of different isotopes) are indicated by colored markings and joined by a dashed line of the same color: hydrogen is indicated in red, magnesium in gray, iron in blue, sodium in violet, nitrogen in green, calcium in cyan, and carbon monoxide in pink. Upper panel:  $^{12}$CO second-overtone band heads seen in the $H$-band spectrum taken in 2018. Lower panel: $^{12}$CO and $^{13}$CO band heads in absorption detected in the $K$-band. The spectra obtained in 2017 (in red) and 2018 (in black) revealed the variability in the strength of the observed bands. Wavelengths are given in angstroms.}
  \label{fig-K-band-variable-CO-bandheads} 
\end{figure}

Figure \ref{fig-K-band-variable-CO-bandheads} (upper panel) displays the second-overtone band heads of $^{12}$CO in absorption from the 2018 $H$-band spectrum. The lower panel compares the $K$-band spectra taken in 2017 (in red) and 2018 (in black), where the variation of the first-overtone band heads of $^{12}$CO is clearly seen. The positions of the $^{13}$CO band heads are also marked and clearly detected in the spectrum from 2017. The 2018 spectrum is too weak to see these \mbox{faint features.}

\begin{figure}[H]   
  \includegraphics[angle=270,width=.67\textwidth]{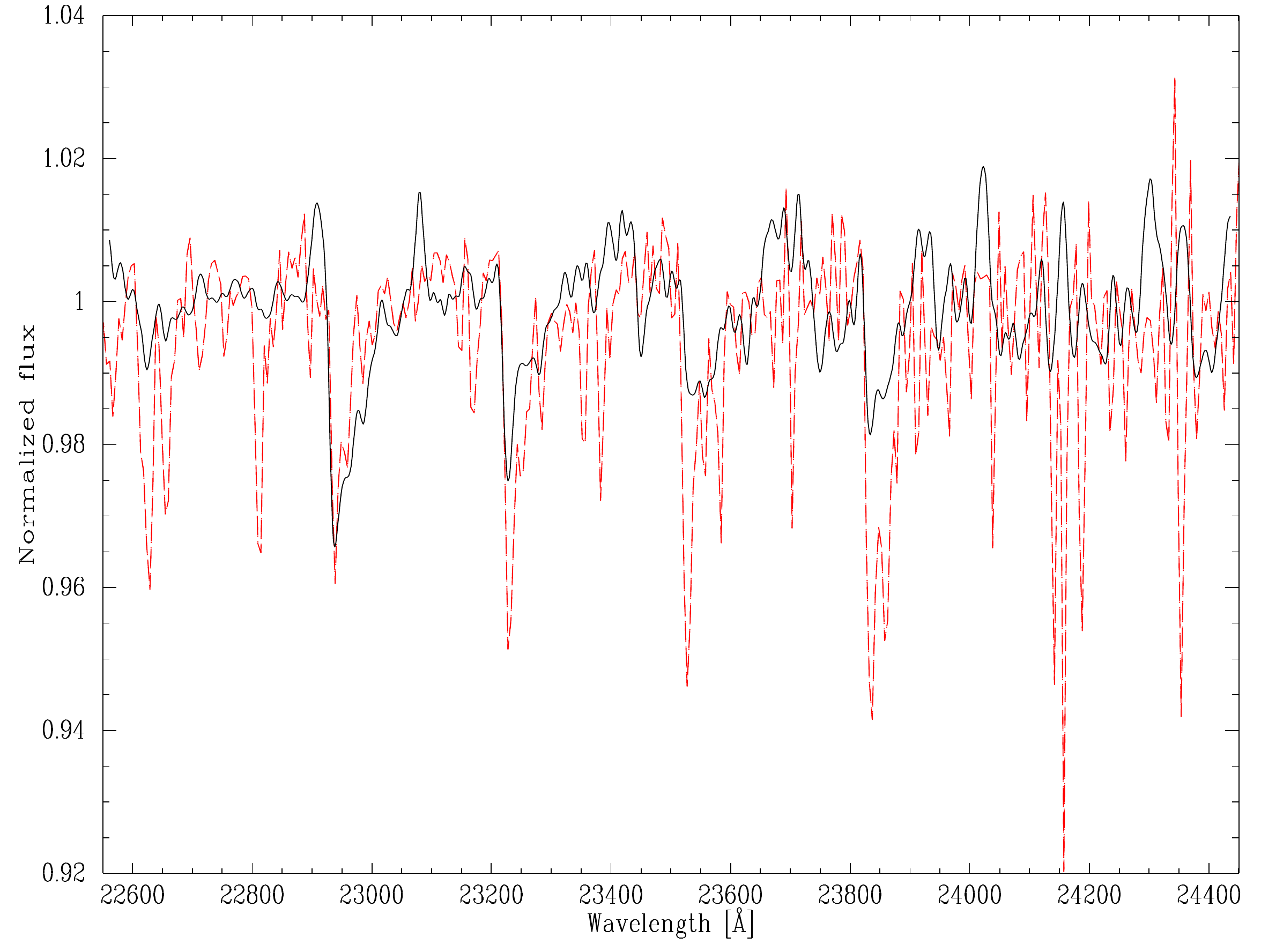}
  \caption{Comparison between the degraded spectrum of MWC\,645 to R$\sim$2\,000 (solid black line) and a G0 Ib-II star, HD 185018 (dashed red line), where the CO(2-0) band head fits well. The other $^{12}$CO absorption bands from MWC\,645 are shallower than those of the template; perhaps they are filled by emission. Wavelengths are given in angstroms. }
  \label{fig-comparison-CO-bands-IRTF-template} 
\end{figure}

The strength of the CO absorption bands in the near-IR spectra of classical late-type stars depends on the stellar effective temperature, T$_{eff}$, and surface gravity, \mbox{$\log$ g \citep{1986ApJS...62..501K, 2008A&A...489..885M}.}\textls[-20]{ The CO absorption becomes deeper when the effective temperature decreases and the luminosity increases. Thus, hot star spectra display no trace of CO features} \mbox{(T$_{eff}\geq$ 5800 K--6000 K,} \citet{1995AJ....110.2415A}), and dwarf stars present weaker CO absorption bands than supergiants. To characterize the cool companion of MWC\,645, responsible for the CO absorption features, and estimate its fundamental parameters, we used the IRTF (NASA Infrared Telescope Facility) Spectral Library \citep{2005ApJ...623.1115C, 2009ApJS..185..289R}, which collects stellar spectra observed with the spectrograph SpeX at a resolving power of R$\sim$2000 and an S/N ratio of about 100 at $\lambda$ < 4 $\upmu$m. We looked for late-type stars with spectral types between F and M and luminosity classes between I and V to compare their spectra with our spectrum from 2017 in the wavelength range from 2.26 to 2.44 $\upmu$m. Figure \ref{fig-comparison-CO-bands-IRTF-template} shows this comparison. The MWC\,645 spectrum (solid black line) was degraded to the resolution of the template spectrum (dashed red line), which corresponds to a G0 Ib-II star (HD\,185018). The intensity of the first $^{12}$CO band head of MWC\,645 coincides reasonably well with that of the early G-type star; however, the rest of the band heads are less intense. The blue edge of the CO(2-0) band head might present an incipient emission. The absorption of the first $^{13}$CO band head is more intense than that displayed by the library star. According to \citet{1997ApJS..111..445W}, the $^{13}$CO isotope is prominent in the supergiants and giants but is not apparent in the dwarfs, although its strength also depends on the initial rotation velocity of the star and the mixing processes that can cause a surface enrichment in $^{13}$C. Otherwise, the absorption lines of neutral metals are less intense than those in the template spectrum, indicating an earlier spectral type (F8-F9 subtypes). Furthermore, the lack of SiO band heads at 4 $\upmu$m, often observed in K0-type stars and later, also points  towards an earlier \mbox{type \citep{2002A&A...394..539H}.}

\citet{2009ApJS..185..186W} presented a spectroscopic library of late spectral-type stellar templates in the $K$-band at a resolving power of R $\sim$ 5900. The authors plotted the equivalent width (EW) of the first CO overtone as a function of T$_{eff}$ (see their Figure~\ref{fig-mwc645-banda-l}) for a stellar sample with T$_{eff}$ in the range 3200--5200 K and different luminosity classes. They measured the EW from the blue edge of the (2-0) band head to the blue edge of the (3-1) band head, more precisely in the window 2.293--2.322 $\upmu$m. We measured the EW of the CO(2,0) band head from the spectrum of 2017 and obtained 2.55 $\pm$ 0.5 \AA. A visual extrapolation of the relation seen in the figure between EW and T$_{eff}$ in the hottest edge of the plot gives an estimation of T$_{eff}$ around 5200 $\pm$ 100 K. From the spectrum obtained in 2018, we measured an EW of the CO(2,0) band head equal to 1.22 $\pm$ 0.1 and estimated a T$_{eff} \sim$ 5300 $\pm$ 100 K.

\vspace{-6pt}

\section{Analysis of the Optical Data}\label{optical}

\subsection{Photometric Light Curve}

\begin{figure}[H]
\includegraphics[width=.63\textwidth,angle=0]{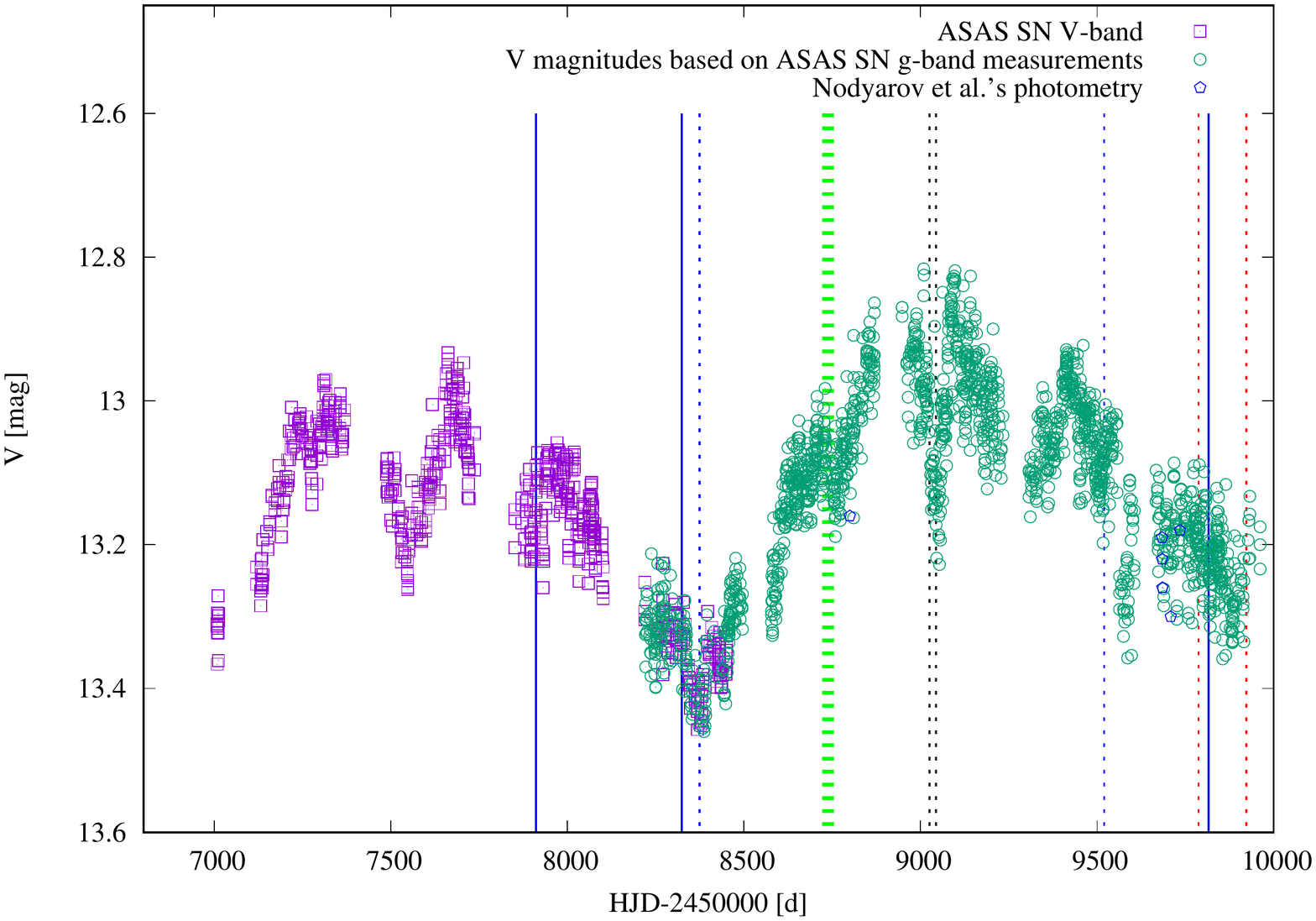}
\caption{Light curve of MWC\,645 from 16 December 2014 up to 11 November 2022 taken from ASAS-SN. The purple squares indicate the {\it V}-band magnitudes, {and the green circles represent the {\it g}-band measurements converted to the {\it V}-band magnitudes.} The conversion has been carried out with the relationship found by \citet{2022ApJ...936..129N}. Their optical photometric observations are also included (blue circles). Vertical solid blue lines mark the dates of our IR observations (2017, 2018, and 2022, respectively); dotted blue lines mark the dates of our optical spectra (2018 and 2021, respectively); and those dotted in green, gray, and red correspond to the spectra downloaded from the BeSS database taken in 2019, 2020, and 2022, respectively. Time is given in heliocentric Julian dates (HJD) minus \mbox{2.45 $\times$ 10$^6$ days.} } \label{fig-asas}
\end{figure}

Figure~\ref{fig-asas} shows the light curve of MWC\,645 taken from ASAS-SN. We applied the relationship derived by \citet{2022ApJ...936..129N} to convert the {\it g}-band magnitudes into {\it V}-band magnitudes. The optical photometry in the {\it V}-band acquired by the authors mentioned above is also included. The dates of the spectroscopic observations presented in this work are marked in the light curve as vertical lines.

The brightness fluctuations of the star up to July 2022 have been reported \linebreak \mbox{by \citet{2022ApJ...936..129N},} who suggested that a new minimum in the light curve might take place in the second half of 2022. As can be seen in the plot, the star continued fading up to the end of October approximately, reaching a minimum of $\sim$0.1 mag brighter than the minimum that occurred in August/September 2018. Then, it began to strengthen in brightness again. 

 \citet{2022ApJ...936..129N} searched for periodicity, excluding visual magnitudes greater than 13.2 mag from their analysis. They derived 69, 145, and 295 days. They attributed the quasi-cyclic photometric variations to variable CS extinction. We applied the Lomb--Scargle method using the IRSA (\url{https://irsa.ipac.caltech.edu/irsaviewer/timeseries} (accessed on 30 November 2022)) time series tool to the {\it V}-band light curve shown in Figure \ref{fig-asas}. We discarded the magnitudes with errors greater than 0.03. The scan of periodic signals with values below ten days gave strong peaks at one day and harmonics of the sidereal day due to the observing cycle. 
 
\begin{figure}[H]
\includegraphics[width=11 cm]{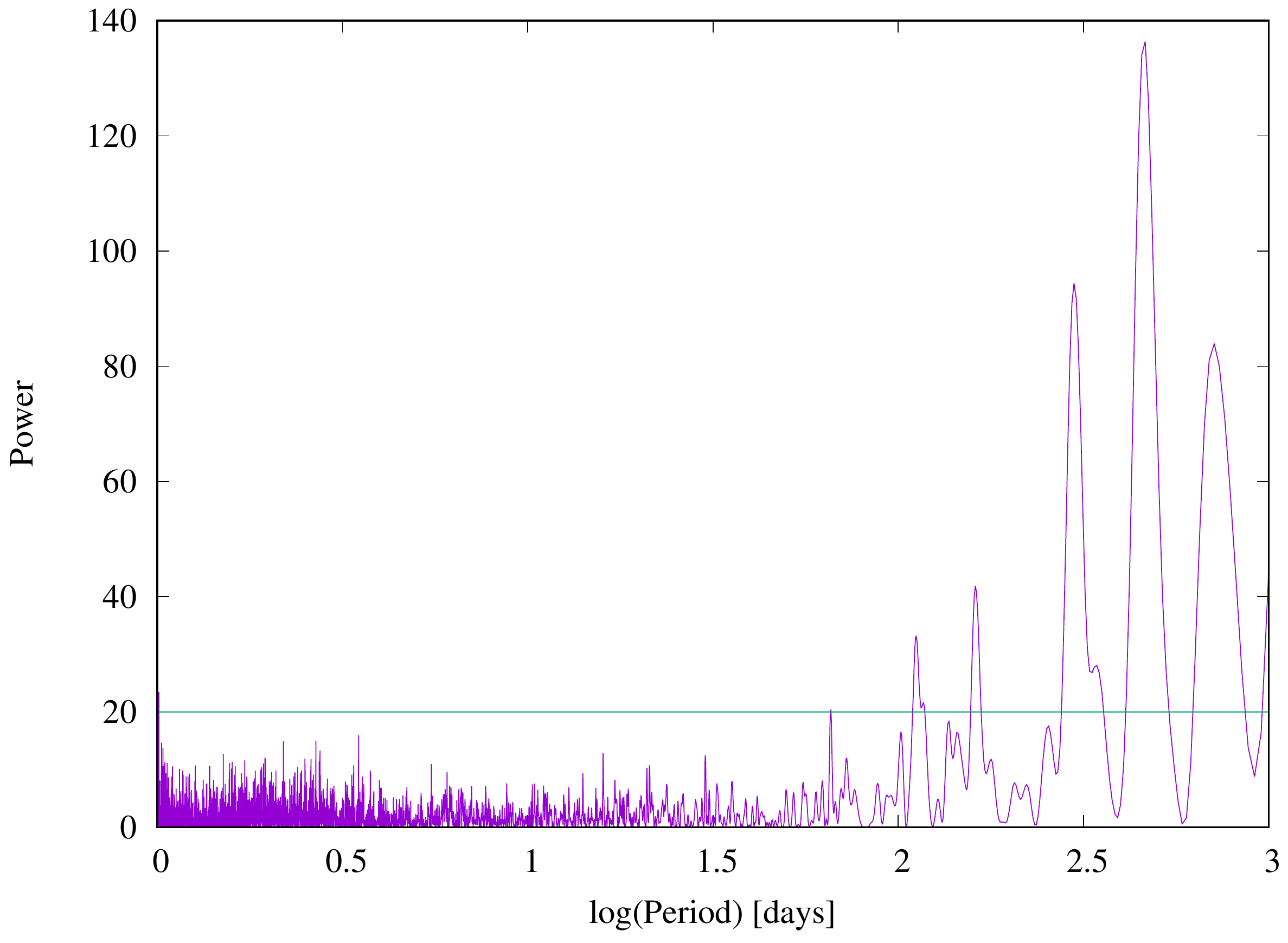}
\caption{Fourier power spectrum of the ASAS-SN light curve of MWC\,645. The periods in days are on a logarithmic scale. The green line shows the confidence level. \label{fig-periodogram}}
\end{figure}

The periodogram for periods greater than one day is shown in Figure~\ref{fig-periodogram}. The \mbox{six peaks} at or above a confidence level of 20 in the power spectrum correspond to periods of approximately 65, 112, 162, 298, 461, and 709 days. We dismissed the last period since the time coverage of the observations is not enough for its precise determination. \mbox{The phase} diagram for each period shows a large scatter of the magnitude points and a small amplitude in their modulation ($\sim$0.2--0.3 mag). We should note that the 65- and 298-day periods were also found by \citet{2022ApJ...936..129N}. As our data spread over a more extended baseline than the one used by the authors mentioned above, this might be a possible explanation for the differences in the other identified periods.

\subsection{Spectroscopic Data}

\begin{figure}[H]
\includegraphics[width=7 cm,angle=-90]{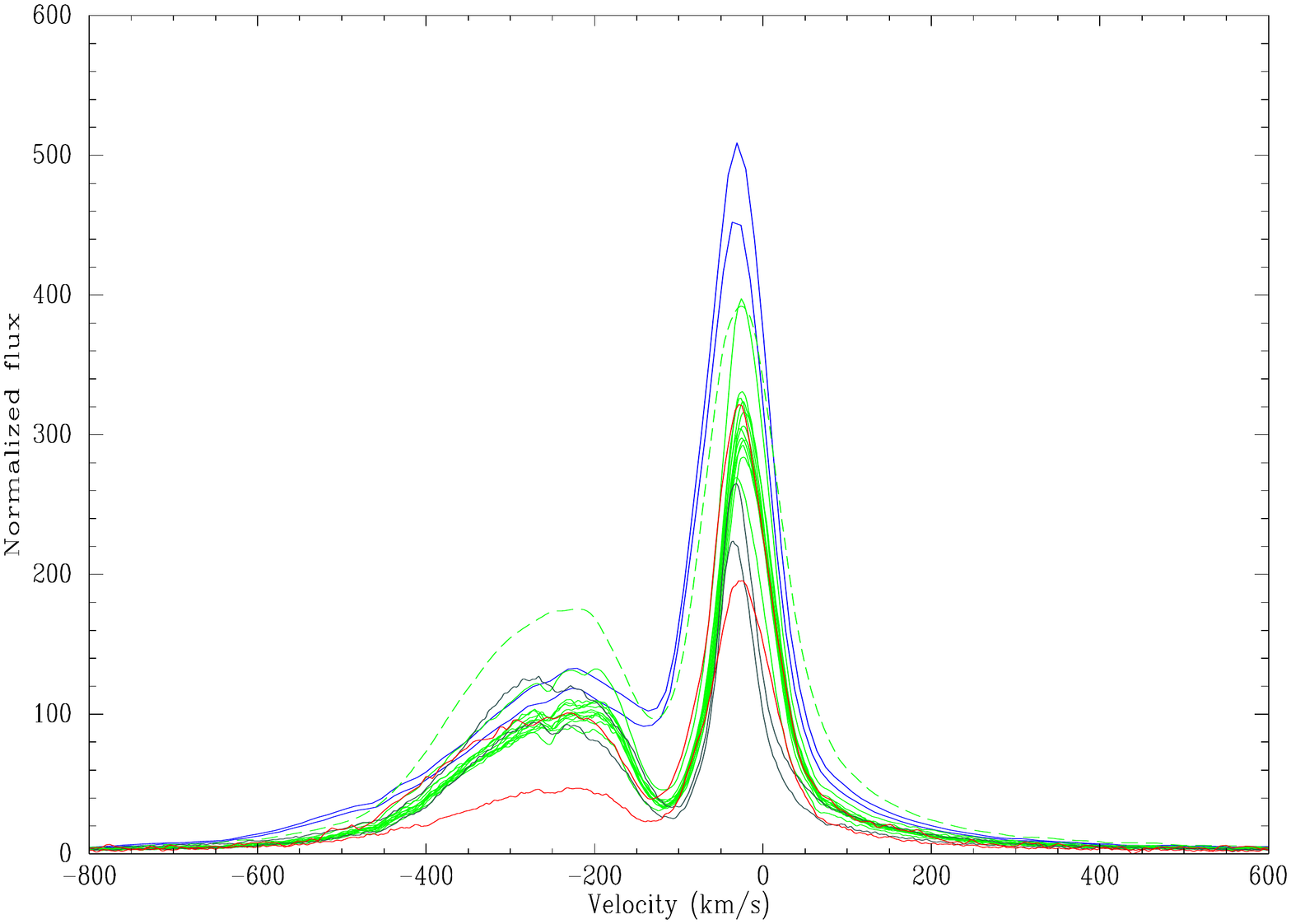}
\caption{H${\alpha}$ line variation of MWC\,645. Spectra taken in 2018, 2019, 2020, and 2022 are displayed in blue, green, gray, and red, respectively. The spectrum with the lowest resolution is plotted with a dashed-line. The heliocentric radial velocity scale is shown in km s$^{-1}$.}\label{fig-halpha}
\end {figure}

The peculiar profile of the H${\alpha}$ line of MWC\,645, composed of a broad blue-shifted peak and a narrow red-shifted one, can be seen in Figure \ref{fig-halpha}, where all the spectra are normalized to the continuum level. This plot shows not only the profile changes over different years (spectra from the same year are displayed in the same color) but also daily. We note that as the BeSS spectra are not corrected by telluric lines, some of the profiles present one or two absorption features superimposed on the blue-shifted emission peak corresponding to water vapor lines of the Earth's atmosphere, which affect the shape of the profile. A variation in the emission strength of both peaks is seen. Using the ’e’ task in the IRAF splot routine, we measured the intensity of the blue (V) and red (R) emission peaks and the total equivalent width of the H$\alpha$ profile, except for the lowest resolution spectrum. Table \ref{tab1} presents these values and the calculated V/R ratios. 
We can see that the V/R ratio presents changes over four years, even doubling its value. We note that the ratio of V/R$\sim$0.3 corresponds to observations close in time (except for one). For this subset, the changes in EW might be mainly due to the continuum-level variations.

The average radial velocity of the blue and red emission components derived from the Ond\v{r}ejov spectra are $-$225 $\pm$ 5 km s$^{-1}$ and $-$31 $\pm$ 3 km s$^{-1}$, respectively, which are in agreement with the values reported by \citet{2003A&A...408..257Z} of $-$218 km s$^{-1}$ and $-$30 km s$^{-1}$ and \citet{2022ApJ...936..129N} of $-$252 $\pm$ 9 km s$^{-1}$ and $-$30 $\pm$ 2 km s$^{-1}$, respectively. Fitting a Gaussian profile to the narrow red component of the H$\alpha$ line, we obtained an average FWHM of 90 $\pm$ 1 km s$^{-1}$. To fit the broad emission component, we built a profile with a red wing symmetrical to the observed blue one and obtained an average FWHM of \mbox{318 $\pm$ 4 km s$^{-1}$. }

Even though the BeSS material is not accurate enough to measure radial velocities, we have estimated them from the different spectra for both emission peaks fitting the components with Gaussian profiles. The average value is $-$229 km s$^{-1}$ and $-$26 km s$^{-1}$ for the blue and red emission peaks, respectively. In Figure \ref{fig-halpha}, a variation in the central wavelength of the H$\alpha$ red emission peak (which is not distorted by telluric lines) can be observed from the different spectra; however, as the wavelength calibration of the BeSS spectra is not well suited for radial velocity determination, we cannot confirm if this change is real. The average FWHMs of the blue and red components are 256 $\pm$ 4 km s$^{-1}$ and 80 $\pm$ 2 km s$^{-1}$, respectively. The H$\alpha$ broad component line profile from the BeSS spectra has a smaller average FWHM than the Ond\v{r}ejov spectra. In the latter, the broad component presents a blue wing with a gentler slope that is also outlined for the red wing, giving a greater width.

\vspace{-10pt} 
\begin{table}[H]
\caption{H$\alpha$ line parameters of MWC\,645. Column 1 indicates the observing date; column 2 the heliocentric Julian date (minus 2.45 $\times$ 10$^6$ d); column 3 the observatory/database where the spectrum was obtained; columns 4 and 5 the emission intensities of the blue and red peaks (V and R) in continuum units, respectively; column 6 the emission intensity ratio of both peaks (V/R); and column 7 the total equivalent width (EW) in \AA. The measurement errors are on the order of 1\% for the intensities and 10\% for the EW values.\label{tab1}}
\newcolumntype{C}{>{\centering\arraybackslash}X}

\begin{adjustwidth}{-\extralength}{0cm}

\begin{tabularx}{\fulllength}{CCCCCCC}
\toprule
\textbf{Obs. Date}	& \textbf{HJD-2450000} & \textbf{Observatory} &    \textbf{V}     & \textbf{R} &   \textbf{V/R}      &  \textbf{EW} \\
\textbf{(yyyy-mm-dd)}	&  &                      &                   &            &                           & \textbf{[\AA]} \\
\midrule
2018-09-11          & 8373.3290   & Ond\v{r}ejov &             118.4                & 452.0         &  0.26 &$-$1980.3\\
2018-09-12          & 8374.2823   & Ond\v{r}ejov &             132.8                & 509.0         &  0.26  &$-$2122.8\\
2019-08-28          & 8724.3870   & BeSS \textsuperscript{1} &   90.2                & 268.9         & 0.33 &$-$1132.2\\  
2019-08-30          & 8726.4279   & BeSS \textsuperscript{2} &   75.1                & 391.9         & 0.45 &$-$2130.7\\
2019-09-01          & 8728.4429   & BeSS &                      95.7                & 283.9         & 0.34 &$-$1237.2\\
2019-09-02          & 8729.4419   & BeSS &                     105.7                & 316.0         & 0.33 &$-$1291.7\\
2019-09-03          & 8730.4365   & BeSS &                     101.5                & 306.1         & 0.33 &$-$1353.7\\
2019-09-05          & 8732.4433   & BeSS &                     100.1                & 292.3         & 0.34 &$-$1147.3\\ 
2019-09-07          & 8734.3942   & BeSS &                     109.7                & 323.5         & 0.34 &$-$1372.1\\
2019-09-10          & 8737.3883   & BeSS &                      99.8                & 296.0         & 0.34 &$-$1292.5\\
2019-09-11          & 8738.3925   & BeSS &                     132.3                & 397.1         & 0.33 &$-$1642.3\\
2019-09-14          & 8741.4241   & BeSS &                     110.5                & 330.9         & 0.33 &$-$1348.1\\
2019-09-19          & 8746.3989   & BeSS &                     100.5                & 297.7         & 0.34 &$-$1129.5\\
2019-09-20          & 8747.3862   & BeSS &                     103.1                & 304.3         & 0.34 &$-$1264.6\\
2019-09-24          & 8751.3722   & BeSS &                     109.1                & 326.1         & 0.33 &$-$1283.5\\
2020-06-25          & 9025.5535   & BeSS &                      94.1                & 223.7         & 0.42 &$-$933.8\\
2020-07-12          & 9043.5126   & BeSS &                     127.4                & 264.8         & 0.48 &$-$1216.8\\
2021-11-01          & 9520.3161   & Tartu &                     ---                 &  ---          & ---    & ---\\
2022-07-26          & 9787.5322   & BeSS &                      47.3                & 195.6         & 0.24 &$-$718.2\\
2022-12-09          & 9923.2775   & BeSS &                     100.5                & 321.2         & 0.31 &$-$1392.3 \\
\bottomrule
\end{tabularx}

\end{adjustwidth}
\noindent{\footnotesize{\textsuperscript{1} For details about the instruments and observers, please visit the web page BeSS database (\url{http://basebe.obspm.fr/basebe/})  (accessed on 2 February 2023).}}  \noindent{\footnotesize{\textsuperscript{2} Spectrum with the lowest resolution.}}
\end{table}

\begin{figure}[H]
\includegraphics[width=6.5 cm,angle=-90]{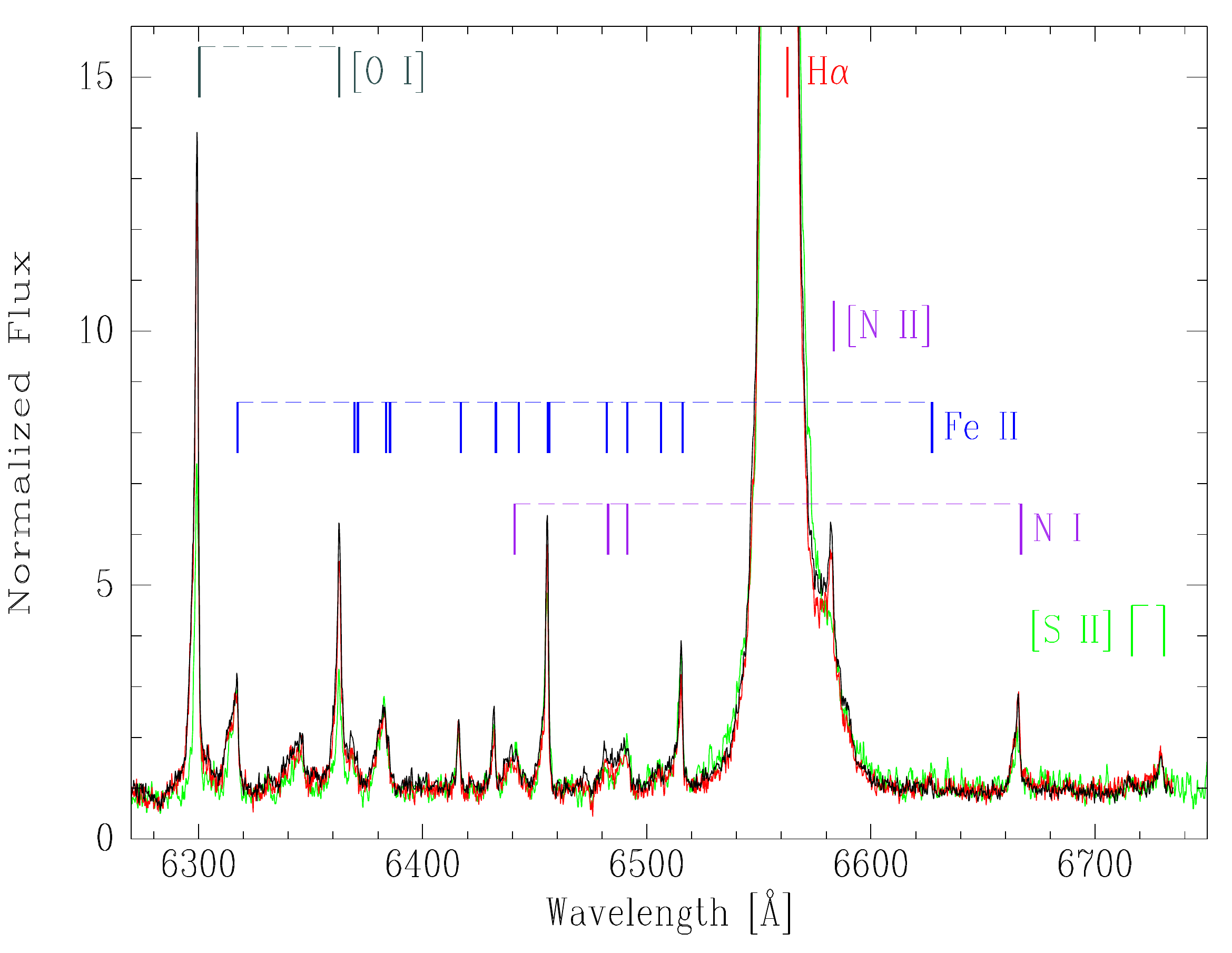}\caption{Example of the emission lines in the surroundings of the H$\alpha$ line of MWC\,645. Ond\v{r}ejov normalized spectra taken in 2018 on 12 September (in red line) and 13 September (in black line) and the 2019 low-resolution spectrum (in green) are shown with the main lines identified by colored markings. The spectral features of a given element (either permitted or forbidden and of different ionization states) are joined by a dashed line of the same color: hydrogen is indicated in red, oxygen in gray, iron in blue, nitrogen in violet, and sulfur in green. Wavelengths are given in angstroms.}\label{fig-oi-lines}
\end{figure}

\newpage
Apart from the H$\alpha$ line, our medium-resolution spectroscopic observations and the 2019 low-resolution BeSS spectrum also show the lines of [O I] $\lambda\lambda$ 6300, 6364 \AA\,. These lines appear single-peaked, although asymmetric, as was previously mentioned \mbox{by \citet{2003A&A...408..257Z}, }which might suggest that the lines are composed of two blended components (see Figure~\ref{fig-oi-lines}). Several permitted Fe II lines and the forbidden lines of [N II] $\lambda$ 6583 \AA\, and [S II] $\lambda\lambda$ 6716, 6731 \AA\, are apparent. We calculated the average heliocentric radial velocity of the emission lines of the Ond\v{r}ejov spectra and the standard error of the mean, obtaining $-$43 $\pm$ 2 km s$^{-1}$. \citet{1996A&AS..120...99J} derived $-$76 $\pm$ 5 km s$^{-1}$, \mbox{and \citet{2022ApJ...936..129N}} found $-$61 $\pm$ 4.3 km s$^{-1}$, which indicates a variation in radial velocity. The He I $\lambda$ 6678 \AA\, transition is absent, as in the spectra studied by the last-mentioned authors \mbox{and \citet{2003A&A...408..257Z}.}
\vspace{-5pt}

\subsection{Global Properties of the Circumstellar Material}

\begin{figure}[H]
\includegraphics[width=.55\textwidth,angle=0]{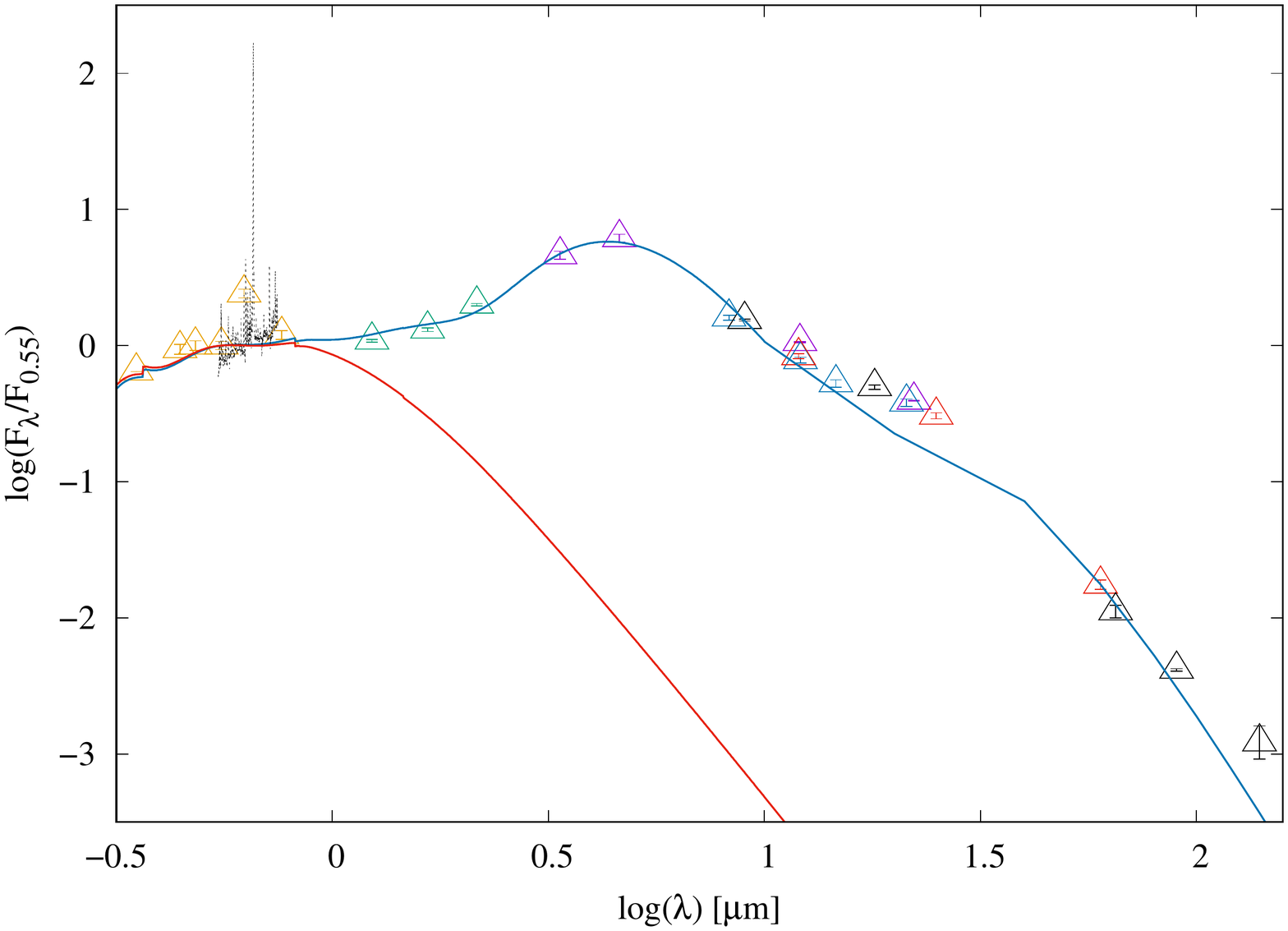}
\caption{Spectral energy distribution of MWC\,645. The open triangles represent the observed photometric data: optical bands (yellow), 2MASS (green) \citep{2006AJ....131.1163S}, WISE (violet) \citep{2010AJ....140.1868W}, MSX (light \mbox{blue) \citep{2003AAS...203.5708E},} IRAS (red) \citep{1984Sci...224...14N}, and AKARI (black) \citep{2010cosp...38.2496Y}. The error bars of the photometric data are included (in most photometric bands, they fall inside the symbols). The low-resolution spectrum acquired in 2021 over the H$\alpha$ region is also displayed (with a dashed black line). The solid red line shows the SED modeled considering the contribution of the photospheric fluxes from both stars, the thermal emission from a gaseous shell close to the system and the effect of the interstellar medium extinction. The solid blue line shows our best-fitting theoretical SED, obtained by adding to the SED plotted in red the contribution of three dusty shells surrounding the stellar system. The flux is normalized to that at $\lambda_{ref}$ = 0.55 $\upmu$m and displayed on a logarithmic scale. The wavelengths are in microns.} \label{sed}
\end{figure}

Figure \ref{sed} shows the spectral energy distribution (SED) of MWC\,645, built from the photometry publicly available from the ultraviolet to the far-IR (0.3 $\upmu$m--140 $\upmu$m). \mbox{The low}-resolution spectrum taken in 2021 in the spectral region of the H$\alpha$ line is also included.

To derive the global physical properties of the CS material, we considered a simple model presented by \citet{2013BAAA...56..163M} and \citet{2018PASP..130k4201A}. The numerical code allows for obtaining the SED assembled by different envelope components. The model assumes the presence of a spherical envelope composed of gas close to the star ($\leq5\, R_{*}$) and (or) dust further away from it ($\geq100\, R_{*}$) \citep{1998ASSL..233...27Z,2010BAAA...53..123M}. The emergent flux is computed from the central star and the envelope (considering that the latter can be reduced to an equivalent shell), applying a plane-parallel solution for the transfer equation. The optical depth $\tau_{\lambda}^G$ and the source function characterize the gaseous shell, which can be described by adding as free parameters the electron temperature T$_G$ and the effective radius R$_G$. \mbox{The dusty} region is treated using an analogous scheme with similar parameters describing the shell: an optical depth $\tau_{\lambda}^D$, a temperature T$_D$, and an effective radius R$_D$. The model allows several dust shell components to be added. The interstellar extinction is also included by an optical depth $\tau_\lambda^{ISM}$. The absorption A($\lambda$) is related to each optical depth through the expression $\tau$ = 0.4\,$\ln(10)$\,A($\lambda$). Using the law given by \citet{1989ApJ...345..245C}, it can be written as \linebreak A($\lambda$)= [R$_V$\,a(1/$\lambda)+b(1/\lambda)$]\,E(B-V), where R$_V$ is the total to selective extinction and E(B-V) is the color excess. We took R$_V^{ISM}$ = 3.1 for the interstellar dust and tried different values of R$_V^D$ greater than 3.1 for the CS dust shell components. The temperature of the dust grains depends on the stellar radiation and the distance from the star center \citep{1999isw..book.....L}. That is, T$_D$(r) = T$_{eff}$\,W(r)$^{[1/(4+p)]}$, where W(r) is the geometrical dilution factor. The parameter p depends on the nature of the dust, but it is usually on the order of one. Furthermore, as the equilibrium temperature should be lower than the dust condensation temperature (typically around 1500 K) to allow the formation of grains, it constrains the distance where condensation can occur, e.g., for a T$_{eff}$ of 18,000 K, the condensation distance is about \mbox{249 R$_*$.}

The code gives the observed flux normalized to that at a reference wavelength $\lambda_{ref}$; we chose $\lambda_{ref}$ = 0.55 $\upmu$m. We assumed that the flux of the central object results from the contribution of both stellar fluxes, for which we considered the Kurucz (\citep{1979ApJS...40....1K}) atmosphere models. Regarding what we know about the stars, we selected models between 17,000 and 20,000 K for the hot binary component and between 4500 and 6000 K for the cool one and explored the flux contribution of the hot and cool components to the total flux in the range of 70\%--90\% and 30\%--10 \%, respectively (similar percentages were suggested \mbox{by \citet{2022ApJ...936..129N}).} Figure \ref{sed} displays our best fit of the theoretical SED (solid blue line) to the observed data. The best-fitting model was computed by taking the photospheric fluxes from a star with T$_{eff}$ = 18,000 K, $\log$\, g = 4.0, and R$_*$ = 3.73 R$_\odot$ and a cool star with \mbox{T$_{eff}$ = 5000 K} and \mbox{$\log$\, g = 1.5.} The contribution of each stellar flux to the total flux is 80\% and 20\% for the hot and cold components, respectively. The resulting envelope has one gaseous shell at \mbox{R$_G$ = 1.15 R$_{\odot}$} with T$_G$ = 16780 K and $\tau_V$ = 0.1. The dusty region comprises three different shells with the following parameters: R$_D^1$ = 348 R$_*$, \mbox{T$_D^1$ = 1310 K,} R$_D^2$ = 3750 R$_*$, \mbox{T$_D^2$ = 507 K, } \mbox{R$_D^3$ = 0.02 pc,} T$_D^3$ = 98 K. The computed total CS visual absorption is A$_V^D$ = 0.097 mag. We obtained a color excess due to the interstellar medium of \mbox{E(B-V)$^{ISM}$ = 0.98 $\pm$ 0.02 mag,} which results in a total visual absorption \mbox{A$_V$ = 3.13 $\pm$ 0.11 mag.} This value agrees with the one derived by \citet{2022ApJ...936..129N}. A disagreement between the theoretical and observed SEDs in the region of $\log\,\lambda$ = 1.0--1.5 is observed. We should note that the employed code can model the thermal emission of the dust, but it cannot address the computation of silicate bands. Thus, the presence of silicate particles might be responsible for the observed difference between the SEDs at 10 $\upmu$m and 18 $\upmu$m. In fact, \citet{2022ApJ...936..129N} have already reported weak emission bumps at these wavelengths.
\vspace{-5pt}
\unskip

\section{Discussion}\label{discussion}

The forest of emission features in the IR spectral ranges presented in this work account for the presence of large amounts of CS gas embedding and veiling MWC\,645, which makes it challenging to characterize the components of the binary system. In addition, the existence of dust revealed by the strong IR excess of radiation above the photospheric fluxes is another element to consider when building a probable scenario.

In previous studies, the hot component of the system was considered to be a B-type \mbox{star \citep{1973ApL....14...65S, 1996A&AS..120...99J},} until \citet{2022ApJ...936..129N} assigned it as an early-B subtype. This assignment was mainly constrained by the absence of He II lines and the possible presence of He I absorption lines suggested by \citet{1996A&AS..120...99J} (see Section \ref{intro}) in the optical spectral range and the identification of the He I lines in emission in the near-IR. We were not able to detect neutral helium lines in either emission or absorption in our data, at least with a signal intensity above the noise level. However, we cannot discard a possible blend of the He I line at 1.083 $\upmu$m (the transition of He I with the highest theoretical intensity that falls in our observed spectral ranges) with a group of Fe II emission lines precisely identified in the same spectral region. We detected Mg II lines in emission at 2.138 $\upmu$m and 2.144 $\upmu$m. According to the works of \citet{2000A&AS..141...65C} and \citet{2001A&A...371..643S}, who studied a representative sample of Be stars in the $H$- and $K$-bands, no evidence of He I features and the simultaneous presence of Br$\gamma$ and Mg II lines in emission indicate a spectral type between B2 and B4. If He I lines are present, the spectral type is B3 or earlier. New high-resolution IR spectra could be valuable to clarify the presence of He I lines. Nevertheless, we cannot discard any variability in neutral helium lines.

The $^{12}$CO absorption bands detected for the first time at 1.62 $\upmu$m and 2.3 $\upmu$m allow us to constrain the spectral type and effective temperature of the cool binary component. \mbox{The spectral} type derived by the best fit of the first $^{12}$CO band head with that of a G0-type star does not agree with the strength of metallic lines, which are seen very weakly in the spectra of MWC\,645. The effective temperature associated with this spectral type for supergiant/giant stars is around 5600 K \citep{1980ARA&A..18..115P,2007MNRAS.378..617K}. Using the EW of the CO(2,0) band head, we determined a T$_{eff}$ average value of about 5\,250 K. However, since the SED of the star in the near-IR has a significant contribution from the hot companion and from the ionized envelope (free-free emission), all spectral features from the cool companion are reduced in their intensity (due to the false continuum). Thus, the temperature determination should be interpreted as an upper limit to the effective temperature.

On the other hand, we might consider that instead of tracing a cool companion, the detected CO absorption might also uncover some mass-ejection episode, such as can be seen in the yellow hypergiant $\rho$ Cas \citep{2006ApJ...651.1130G} or in the eruptive variable V838 Mon \citep{2007A&A...467..269G}, when molecular absorption bands start to develop in the spectrum while the star's brightness fades. In the case of $\rho$ Cas, the CO bands turn from absorption into emission when the star reaches the next maximum brightness phase in which it is too hot for molecule condensation in its extended outer layers \citep{2019MNRAS.483.3792K}. This finding agrees with the scenario that mass was ejected into the environment, radiating while it is expanding and cooling. Since we have only two observations, we cannot trace the evolution of the molecular absorption along the light curve. However, it is interesting to mention that the most intense CO absorption (seen in 2017) occurs when the brightness begins to increase after a seeming local minimum. This observation presents an emission peak blueward of the CO(2,0) absorption band head that is blue-shifted at about 308 km s$^{-1}$. If it is indeed due to CO, it might suggest a high-velocity molecular outflow. Furthermore, the considerably weaker CO absorption spectrum observed in 2018 might be interpreted as partially filled with circumstellar emission. Moreover, the absorption-line spectrum ascribed to the cool companion might have originated in an optically thick disk. \citet{2018A&A...617A..79P} suggested that the absorption features seen in the spectrum of the FS\,CMa star, MWC\,623, are formed in an equatorial disk viewed nearly edge-on, which acts as a pseudo-photosphere.

When we look at the light curve, we see global qualitative similarities between the range mostly recorded in {\it V}-band magnitudes up to the deepest minimum (which occurred in August/September 2018) and that traced by the {\it g}-band magnitudes from this minimum up to January 2023. In both time lapses of about four years, we can distinguish a well-outlined dip (at HJD-2450000 $\sim$ 7500 d and $\sim$ 9000 d) less intense than the main minimum. Although they present slightly different shapes and depths, and the light curve afterwards reaches a different maximum magnitude level, they are alike. This similar pattern suggests that the dominant source of variability is the same. Variable CS (or circumbinary) extinction along the line of sight due to dust clumps might be responsible for these photometric variations \citep{1997ApJ...475L..41M,1999AJ....118.1043H,2003A&A...409..169B}. Our simple model to fit the SED allowed us to derive the global properties of the dusty envelope. Despite its spherical geometry, it traces different components of optically thin dust, the first located at a distance of $\sim$ 6 AU. This distance agrees with the innermost dusty disk radius of $\sim$ 5 AU derived for the star FS\,CMa through aperture-synthesis imaging in the $L$ and $N$ bands \citep{2022A&A...658A..81H}. Material orbiting the system at this distance at Keplerian velocity would have a period of $\sim$ 5 years. A warped inner edge of the disk can also produce variable extinction \citep{2013MNRAS.429.1213B}.

Previous studies of MWC\,645 have not reported variations in the V/R ratio of the H$\alpha$ line. Observations made over the last 30 years, although discontinuous in time, have not revealed an inversion in the intensity ratio of the peaks, always showing V/R < 1. We noticed that the V/R ratio derived from our spectra varies from 0.2 to 0.5 (see Table \ref{tab1}). We also calculated the V/R ratio from the intensities of both H$\alpha$ peaks included in the work of \citet{2022ApJ...936..129N}. Their observations are from 2004 to 2021 and not continuous. We found a variable V/R ratio with values between 0.3 and 0.9. The only two spectra from dates included in the light curve range, October 2016 (HJD-2450000 $\sim$ 7680 d) and November 2021 (HJD-2450000 $\sim$ 9\,545 d), present a V/R ratio of 0.7 and 0.3, respectively. Spectroscopic monitoring of the H$\alpha$ line could help scan and characterize this variability and search for any periodicity in the changes of the peak intensities that could be related to the rotation of a density perturbation in the disk \citep{1991PASJ...43...75O} or an orbital motion \citep{1975BAICz..26...65K}. Zickgraf computed line profiles assuming a latitude-dependent wind model with a dust disk and obtained similar line profiles to the observed H$\alpha$ profile with V/R < 1 for an intermediate inclination angle.

The dimming in optical brightness occurs over a long time, and, along this phase, the H$\alpha$ emission changes. The V/R ratio decreases, and the EW increases (with the smallest value of the V/R relation and the maximum EW at the minimum of the light curve). This fact seems to be associated with a change in the amount of the circumstellar material, \mbox{and not} only due to a natural increase in emission intensity during the light curve minimum. Also, the FWHM of the blue emission component is the largest at this point. This H$\alpha$ line strengthening might be attributed to enhanced mass loss (or a mass ejection episode).  Similar variations in the emission of the spectral lines and brightness have been observed in the FS\,CMa star, MWC\,728. The light curve of MWC\,645 suggests a (possible) periodic behavior. If this were true, the H$\alpha$ enhancement might result from a mass transfer process during periastron passage in an eccentric binary system, with a period of the order of 4 years. \mbox{A denser} observational grid with high-quality spectra is needed to study the link between mass loss and brightness behavior.

A detailed calculation of the H$\alpha$ line profile based on a physically consistent model is a difficult task. However, it would be valuable to explore models with simplified assumptions to gain insight into the system geometry and the structure of the CS \mbox{matter \citep{2010ApJ...721.1079C,2018A&A...617A..79P}.} More complex scenarios considering non-conservative mass transfer between the binary components should be considered to draw a picture of the structures involved in the emission processes \citep{2015A&A...577A..55D, 2021A&A...645A..51B}. 
\newpage
\section{Conclusions}\label{conclusions}

In this paper, we have studied the FS\,CMa-type object, MWC\,645, a recently confirmed binary system. We have presented IR medium-resolution spectra covering the $J$-, $H$-, $K$-, and $L$-bands and identified the main spectral features. We have reported the presence of CO bands in absorption for the first time. We have searched for periodicity in the light curve and a possible correlation between its behavior and the spectroscopic optical data. We found that the photometric variations could be explained by variable extinction along the line of sight. In addition, we noted that the stellar brightness fading is accompanied by the enhancement of the H$\alpha$ line emission, which might be due to mass ejection events. Finally, a proper fitting to the observed SED was found, giving a global picture of the gaseous and dusty structures that could enshroud the binary.

 Simultaneous optical and near-IR spectroscopy during the following brightness minimum would be very useful for tracing the onset and progress of the possible mass transfer. Such an understanding is utmost for deepening the comprehension of binary evolution in general and of the nature of this fascinating object in particular.

\vspace{6pt} 



\authorcontributions{Conceptualization, A.F.T., M.L.A., and M.K.; methodology, A.F.T., M.L.A., and M.K.; software, A.F.T., M.L.A., and L.V.M.; formal analysis, A.F.T. and M.L.A.; investigation, A.F.T., M.L.A., M.K., L.V.M., and T.E.; resources, A.F.T., M.L.A., and M.K.; data curation, A.F.T., M.L.A., M.K., L.V.M., and T.E.; writing---original draft preparation, review and editing,  A.F.T., M.L.A., M.K., L.V.M., and T.E.; visualization, A.F.T., M.L.A., and M.K.; funding acquisition, A.F.T., M.L.A., M.K., and T.E. All authors have read and agreed to the published version of the manuscript.}

\funding{A.F.T. and M.L.A. acknowledge financial support from the Universidad Nacional de La Plata (Programa de Incentivos 11/G160) and CONICET (PIP 1337), Argentina. M.K. acknowledges financial support from the Czech Science Foundation (GA\v{C}R, grant number 20-00150S). The Astronomical Institute of the Czech Academy of Sciences, Ond\v{r}ejov, is supported by project RVO: 67985815. This project has received funding from the European Union’s Framework Programme for Research and Innovation Horizon 2020 (2014-2020) under the Marie Skłodowska-Curie Grant Agreement No. 823734. T.E. gratefully acknowledges financial support from the Estonian Ministry of Education and Research through the Estonian Research Council institutional research funding IUT40-1, from the European Union European Regional Development Fund project KOMEET 2014-2020.4.01.16-0029.}

\institutionalreview{Not applicable.}

\informedconsent{Not applicable.}

\dataavailability{The data involved in this research are available on request from \mbox{the authors.}} 

\acknowledgments{The authors are grateful to the referees, whose comments and suggestions helped to improve the paper. This work has made use of IRAF, which is distributed by the National Optical Astronomy Observatory, operated by the Association of Universities for Research in Astronomy (AURA) under a cooperative agreement with the National Science Foundation; the BeSS database operated at LESIA, Observatoire de Meudon, France (\url{http://basebe.obspm.fr} (accessed on 2 February 2023)); the SIMBAD database and the VizieR catalog access tool, both operated at CDS, Strasbourg, France; the NASA Astrophysics Data System (ADS); the NASA IRTF (Infrared Telescope Facility) Spectral Library; and the NASA IRSA period search tool and the All-Sky Automated Survey for Supernovae (ASAS-SN). This paper is based on observations obtained at (i) Ond\v{r}ejov Observatory (Czech Republic) with the Perek 2 m telescope; (ii) Tartu Observatory (Estonia); and (iii) the international Gemini Observatory, a program of NSF’s NOIRLab, which is managed by the Association of Universities for Research in Astronomy (AURA) under a cooperative agreement with the National Science Foundation on behalf of the Gemini Observatory partnership: the National Science Foundation (United States), National Research Council (Canada), Agencia Nacional de Investigaci\'{o}n y Desarrollo (Chile), Ministerio de Ciencia, Tecnolog\'{i}a e Innovaci\'{o}n (Argentina), Minist\'{e}rio da Ci\^{e}ncia, Tecnologia, Inova\c{c}\~{o}es e Comunica\c{c}\~{o}es (Brazil), and Korea Astronomy and Space Science Institute (Republic of Korea) under program IDs GN-2017A-Q-62, GN-2018A-Q-406, and GN-2022B-Q-225. This work has made use of the ground-based research infrastructure of Tartu Observatory, funded through the projects TT8 (Estonian Research Council) and KosEST (EU Regional Development Fund). The authors thank L.S. Cidale for many \mbox{fruitful discussions.}}

\conflictsofinterest{The authors declare no conflict of interest.} 

\newpage
\abbreviations{Abbreviations}{
The following abbreviations are used in this manuscript:\\

\noindent 
\begin{tabular}{@{}ll}
CS & Circumstellar\\
CCD & Charged-coupled device\\
EW & Equivalent width\\
V & Blue peak intensity\\
R & Red peak intensity\\
V/R & Blue-to-red emission peak ratio\\
FWHM & Full width at half maximum\\
HJD & Heliocentric Julian date\\
IR & Infrared\\
SED & Spectral energy distribution\\
\end{tabular}
}




\begin{adjustwidth}{-\extralength}{0cm}

\reftitle{References}

\PublishersNote{}
\end{adjustwidth}
\end{document}